\documentclass{kluwer}
\usepackage{graphicx} 
\begin{document}
\bibliographystyle{klunamed}
\begin{article}
\begin{opening}
\title{Preprocessing of vector magnetograph data for a non-linear force-free
magnetic field reconstruction.}
\author{T. \surname{Wiegelmann}\email{wiegelmann@linmpi.mpg.de}}
\author{B. \surname{Inhester}}
\institute{Max-Planck-Institut f\"ur Sonnensystemforschung,
Max-Planck-Strasse 2, 37191 Katlenburg-Lindau, Germany}
\author{T. \surname{Sakurai}}
\institute{National Astronomical Observatory of Japan, Solar Physics Division, Osawa
2-21-1, Mitaka, Tokyo 181-8588, Japan }


\runningtitle{Preprocessing}
\runningauthor{Wiegelmann et al.}

\begin{ao}
\end{ao}

\begin{motto}

{\large \bf Solar Physics, Vol. 233, 215-232 (2006)}
\end{motto}
\begin{abstract}
 Knowledge regarding the coronal magnetic field is important for
 the understanding of many phenomena, like flares and coronal
 mass ejections. Because of the low plasma beta in the solar corona
 the coronal magnetic field is often assumed to be force-free and we
 use photospheric vector magnetograph data to extrapolate the magnetic
 field into the corona with the help of a non-linear force-free optimization
 code. Unfortunately the measurements of the photospheric magnetic field
 contain inconsistencies and noise. In particular the transversal components
 (say Bx and By) of current  vector magnetographs have their uncertainties.
 Furthermore the magnetic field in the photosphere is not necessary force-free
 and often not consistent with the assumption of a force-free field above. We
 develop a preprocessing procedure to drive the observed non force-free data
 towards suitable boundary conditions for a force-free extrapolation. As a result
 we get a data set which is as close as possible to the measured data and consistent
 with the force-free assumption.

\end{abstract}


\abbreviations{\abbrev{KAP}{Kluwer Academic Publishers};
   \abbrev{compuscript}{Electronically submitted article}}

\nomenclature{\nomen{KAP}{Kluwer Academic Publishers};
   \nomen{compuscript}{Electronically submitted article}}

\classification{JEL codes}{D24, L60, 047}
\end{opening}

\section{Introduction}
\label{sec1}
 The structure of the solar corona out to a few solar radii is dominated
 by the magnetic field. Knowledge regarding the coronal magnetic
 field is therefore important to understand physical processes
 like flares and coronal mass ejections.
 Unfortunately direct measurements
 of the magnetic field are difficult for the following reasons:
 The high plasma temperature in the corona broadens the line profile
 orders of magnitudes above the Zeeman splitting.
 In addition coronal lines are optical thin and consequently the line-of-sight integrated
 character of the measurements complicates their interpretation.

 It has been proposed to use magnetic sensitive coronal line
 observations in order to constrain the coronal magnetic field
 ( see, e.g. \inlinecite{house77} , \inlinecite{arnaud:etal87}, \inlinecite{judge98}).
 \inlinecite{lin:etal04}
 demonstrated that Zeeman-effect
 measurements can be performed. However, due to the line-of-sight effect
 the interpretation of these lines is difficult.
 Also, recent measurements of chromospheric and lower coronal magnetic
 fields were made for a few individual cases, e.g.
 \cite{solanki:etal03,lagg:etal04}. Such multi-wavelength spectropolarimetric
 observations have not been done
routinely in the past but are believed to become much more
commonplace, and reliable, in the near future.

 As an alternative to these direct measurements, methods have been
 developed to use the magnetic field observed on the photosphere for
 an extrapolation into the corona. The extrapolation is not free
 from assumptions regarding the coronal plasma. It is helpful, that
 the low and middle corona contains a low $\beta$ plasma which allows
 to neglect the plasma pressure in first order and use a force-free
 magnetic field model.
 A force-free magnetic field is characterized by
 the electric currents parallel to the magnetic field lines, i.e.,
 \begin{eqnarray}
   \mathbf{\nabla}\times\mathbf{B} &=& \alpha \mathbf{B} \label{curlB=aB} \\
  \mathbf{\nabla}\cdot \mathbf{B} &=& 0
 \end{eqnarray}
 where the coefficient $\alpha$ is constant along the field lines but
 may vary between different field lines. Since (\ref{curlB=aB}) due to
 its intrinsic non-linearity is difficult to solve, simplifications
 of (\ref{curlB=aB}) are sometimes used like
 potential fields ($\alpha$ = 0, i.e., no currents; see e.g.
 \cite{schmidt64,semel67,cuperman:etal89}) and
 linear force-free fields ($\alpha$ = const; see e.g.
 \cite{chiu:etal77,seehafer78,seehafer82,alissandrakis81,%
 semel88,demoulin:etal92}).

 The non-linear force-free case
 \cite{sakurai81,wu:etal90,roumeliotis96,amari:etal97,yan:etal00,%
 wheatland:etal00,wheatland04,regnier:etal02,wiegelmann:etal03,%
 wiegelmann04,valori:etal05} is challenging both theoretically due to
 the non-linearity of the underlying mathematical problem and
 observationally because a measurement of the full magnetic vector
 on the photospheric is required.
 For investigations of instabilities like filament eruptions, it
 is essential to consider the general non-linear force-free field case,
 because with the simplified field models mentioned above the free
 magnetic energy which might drive the instability cannot be described

 Moreover, a comparison between extrapolations and measured fields
 \cite{solanki:etal03} in a newly developed active region revealed
 that a non-linear model agrees better with the observations
 than a potential and linear force-free magnetic field model
 \cite{wiegelmann:etal05}.

 A non-linear force-free extrapolation of coronal magnetic fields from
 measured photospheric data is a challenging problem for several reasons.
 The magnetic field in the photosphere is not necessary force-free
 (see \inlinecite{gary01}) and one would rather prefer to use the vector-magnetic
 field at the basic of the corona as boundary condition for a non-linear
 force-free extrapolation, but here measurements are usually not available.

 Another problem is the $180^o$ ambiguity in the transversal components
 of the photospheric data, which has to be removed with e.g. the minimum
 energy method \cite{metcalf94}. An additional complication is that
 the transversal components are more difficult to measure and bear
 much more noise than the line-of-sight component.
 For the data of the Solar-Flare Telescope (SFT)
 used in section \ref{sec5}, for example,
 the noise level is about $10 G$ for the line-of-sight magnetic field
 and $100 G$ for the transverse field.
 The full disk vector magnetograph SOLIS
 (Synoptic Optical Long-term Investigations of the Sun, U.S. National Solar
 Observatory, Kitt Peak)
 is expected to have a
 noise level of about $1 G$ in the line-of-sight and about $50 G$ in
 the transverse components (C. Keller, private communication).
 The reason for the much higher noise in the transversal magnetic
 field $(B_t)$ compared with the line-of-sight field ($B_L$)
 can be understood by the way how the photospheric magnetic field is
 derived from the four measured Stokes components.
 Roughly $B_L = c1 \, V/I$ where $V$ is the circular polarization and $c1$ is a constant.
 Then
\begin{equation}
\delta B_L = c1 \frac{\delta V}{I}  \propto \frac{\delta  I}{I}  \propto \sqrt{I},
\label{eq3}
\end{equation}
where $\delta I$ is the noise in $I$. This photon noise is independent of $B_L$.
Even if the $Q$ and $U$ signals are generally weaker than the $V$ signal, the major
source of errors in the force balance is the noise in $B_t$:
$B_t ^2= c2 \, \sqrt{Q^2 +U^2}/I$
where $Q$ and $U$ are linear polarization intensities and $c2$ is another constant. Then
\begin{equation}
2 B_t \, \delta B_t = c2 \, \frac{Q \delta Q+U \delta U}{\sqrt{Q^2+U^2} I}
\sim c2 \, \frac{\sqrt{\delta Q^2+ \delta U^2}}{I} \sim c2 \, \frac{\delta I}{I} \sim const.
\label{eq4}
\end{equation}
As a consequence the transvers magnetic field noise is in particular high in
weak field regions. In strong field regions
vector magnetic field measurements based on full Stokes
spectro-polarimetry generally yield uncertainties of a few
degrees in the orientation angles of the field vector in
active regions.  This situation is also constantly improving
with improved instrumentation.

 For the practical computation of the coronal magnetic field it is
 also helpful if the boundary data are smoothed to some degree. Short
 wavelength fluctuations in the surface magnetic field die out rapidly
 with height (for a potential field with $\exp(-kz)$ for scales of
 horizontal wavenumber $k$). Hence, boundary data with scales much
 finer than the height of the numerical box require a very fine grid to
 be sufficiently resolved but they hardly have a effect on the result,
 except in a very small boundary layer above the surface. To keep the
 numerical effort limited, the boundary data are therefore usually
 smoothed.
 In this context it should also be noted that the force-free equation
 (\ref{curlB=aB}) is scale invariant if $\alpha$ is rescaled
 appropriately. Hence, by smoothing of the boundary data we do not
 lose any physics which we have not already cast away by the
 restriction to a stationary force-free field model.

 A more fundamental requirement on the boundary data is its consistency
 with the force-free field approximation. As has been shown by
 \inlinecite{aly89}, a balance of the total momentum and angular momentum
 exerted onto the numerical box by the magnetic field leads to a
 set of boundary integral constraints for the magnetic field. These
 constraints should also be satisfied on the solar surface for the
 field at the coronal base in the vicinity of a sufficiently isolated
 magnetic region and in a situation where there is no rapid
 dynamical development.

 In summary, the boundary data for the force-free extrapolation should
 fulfill the following conditions:
 \begin{enumerate}
 \item The data should coincide with the photospheric observations
   within measurement errors.
 \item The data should be consistent with the assumption of a force-free
  magnetic field above.
 \item For computational reasons (finite differences) the data should be
 sufficiently smooth.
 \end{enumerate}
 Within this work we describe a numerical procedure, written in IDL,
 which pre-processes  vector magnetograph data so that the above conditions
 are satisfied as close as possible.

 We outline the paper as follows. In section \ref{sec2} we specify
 how to check if a given measured vector magnetograms is consistent
 with the assumption of a force-free magnetic field above. We describe,
 how to derive consistent force-free boundary conditions for a force-free
 magnetic field extrapolation from the measured data in section \ref{sec3}.
 In section \ref{sec4} we use a known semi-analytic force-free model to
 check our method and we apply the method in section \ref{sec5} to
 an example of the observed vector magnetogram.
%
\section{Consistency check of vector magnetograms}
 \label{sec2}

 In the following we assume that the $180^{\circ}$ ambiguity in the
 measured transverse components has been removed and the magnetogram
 has been observed close to the disk center. ``Close'' here means that
 the vertical component $B_z$ in our numerical box can be more or less
 identified with the line-of-sight component and bears a smaller
 measurement error than the horizontal components $B_x$ and $B_y$.
 Since, as stated above, the measurement error between the
 line-of-sight and the transverse components differs by about
 one order of magnitude,
 the angle between line-of-sight and vertical should not
 exceed $\simeq$ 0.1 radian.

 Another a-priori assumption about the photospheric data is that the
 magnetic flux from the photosphere is sufficiently distant from the
 boundaries of the observational domain and
 the net flux is in balance, i.e.,
 \begin{equation}
  \int_{S} B_z(x,y,0) \;dx\,dy =0.
 \end{equation}
 Generally, the flux balance criterion has to be applied to the whole,
 closed surface of the numerical box.
 However, we can only measure the magnetic field vector on the bottom
 photospheric boundary and the contributions of the lateral and top
 boundary remain unspecified. However, if a major part of the known flux
 from the bottom boundary is uncompensated, the final force-free magnetic
 field solution will markedly depend on how the uncompensated flux is
 distributed over the other five boundaries.
 This would result in a major uncertainty of the final force free magnetic
 field configuration. We therefore demand the flux balance
 to be satisfied with the bottom data alone. If this is not the case,
 we classify the reconstruction problem as not uniquely solvable within
 the given box.

 \inlinecite{aly89} used 
 the virial theorem to define which conditions a vector magnetogram has
 to fulfill to be consistent with the assumption of a force-free field
 above.
 We repeat here the force-free and torque-free condition
 and refer to the paper of \cite{aly89} for details.
\begin{enumerate}
\item The total force on the boundary vanishes
$$\int_{S} B_x B_z \;dx\,dy = \int_{S} B_y B_z \;dx\,dy =0$$
$$\int_{S} (B_x^2 + B_y^2) \; dx\,dy = \int_{S} B_z^2 \; dx\,dy.$$
\item The total torque on the boundary vanishes
$$\int_{S} x \; (B_x^2 + B_y^2) \; dx\,dy = \int_{S} x \; B_z^2 \; dx\,dy $$
$$\int_{S} y \; (B_x^2 + B_y^2) \; dx\,dy = \int_{S} y \; B_z^2 \; dx\,dy $$
$$\int_{S} y \; B_x B_z \; dx\,dy = \int_{S} x \; B_y B_z \; dx\,dy $$
\end{enumerate}
Note that if condition 1) is fulfilled, the constraints 2) are independent
on where the origin for ${x,y}$ is located. In our code
the origin is usually at the
lower left corner of the bottom boundary face.

As with the flux balance, the Aly integral criteria in general have to be
applied to the whole surface of the numerical box. Since we assumed that
the photospheric flux is sufficiently concentrated in the center and
the net flux is in balance, we can expect the magnetic field on the lateral
and top boundary to remain small and hence these surfaces will not
yield a large contribution to the integrals of the Aly-constraints
above.
We therefore impose the Aly-criteria on the bottom boundary alone.

 \inlinecite{aly89} already pointed out that the magnetic field is probably
 not force-free in the measured region because the plasma $\beta$
 in the photosphere is of the order of one and pressure and gravity
 forces are not negligible.
 We however expect that the observed photospheric field is, after
 removing scales below a super-granular diameter by smoothing,
 representative for the field at the coronal base.
 In the corona, however we have $\beta \approx 10^{-4}$ and consequently
 the magnetic field should be close to force-free in a stationary
 situation.

 To quantify the quality of vector magnetograms with respect to the above
 criteria we introduce three dimensionless parameters:
\begin{enumerate}
\item The flux balance parameter
$$
\epsilon_{\mbox{flux}}=\frac{\int_{S} B_z \;dx\,dy }{\int_{S} | B_z | \;dx\,dy }
$$
\item The force balance parameter $\epsilon_{\mbox{force}}=$

$$
\frac{|\int_{S} B_x B_z \;dx\,dy| + |\int_{S} B_y B_z \;dx\,dy|+
|\int_{S} (B_x^2+B_y^2)-B_z^2 \;dx\,dy |}
{\int_{S} (B_x^2+B_y^2+B_z^2) \;dx\,dy}
$$

\item The torque balance parameter $\epsilon_{\mbox{torque}}=$

$$
\frac{|\int_{S} x ((B_x^2 + B_y^2)-B_z^2)  dx dy|+
|\int_{S} y ((B_x^2 + B_y^2)-B_z^2) \; dx dy|+
|\int_{S} y  B_x B_z-x  B_y B_z dx dy|}
{\int_{S} \sqrt{x^2+y^2} \; (B_x^2+B_y^2+B_z^2) \;dx\,dy}
$$

\end{enumerate}
 An observed vector magnetogram is then flux-balanced and consistent with the
force-free assumption if: $\epsilon_{\mbox{flux}} \ll 1, \;  \epsilon_{\mbox{force}}
\ll 1, \; \epsilon_{\mbox{torque}} \ll 1$.

\section{Method}
\label{sec3}

 Even if we choose a sufficiently flux balanced isolated active region
 $(\epsilon_{\mbox{flux}} \ll 1)$ we find that usually the force-free
 conditions $ \epsilon_{\mbox{force}} \ll 1, \;\epsilon_{\mbox{torque}} \ll 1$
 are not fulfilled for measured vector magnetograms. We conclude therefore,
 that non-linear force-free extrapolation methods may not be used directly
 on observed vector magnetograms, in particular not on the very noisy
 transverse photospheric magnetic field measurements.
 The large noise in the transverse components of the photospheric field vector
($\sim$ the horizontal $B_x$ and $B_y$ at the bottom boundary)
 gives us the freedom to adjust these data within the noise level.
 We use this freedom to drive the data towards being more
 consistent with Aly's force-free and torque-free conditions.

 As a measure, how well a photospheric magnetic field agrees with
 Aly's criteria, the observed data and the smoothness condition,
 we define the following functional which adds up the chi-square
 deviations from all individual constraints:
 \begin{equation}
  L = \mu_1 L_1 + \mu_2 L_2 + \mu_3 L_3 + \mu_4 L_4
 \end{equation}
where
\begin{eqnarray}
L_1 &=& \left[ \left(\sum_p B_x B_z \right)^2
              +\left(\sum_p B_y B_z \right)^2
              +\left(\sum_p B_z^2-B_x^2-B_y^2 \right)^2
        \right] \nonumber \\
L_2 &=& \left[ \left(\sum_p x  \left(B_z^2-B_x^2-B_y^2 \right) \right)^2
              +\left(\sum_p y  \left(B_z^2-B_x^2-B_y^2 \right) \right)^2
        \right. \nonumber \\ & & \left. \hspace*{0.8em}
              +\left(\sum_p y B_x B_z -x B_y B_z \right)^2
        \right] \nonumber \\
L_3 &=& \left[ \sum_p \left(B_x-B_{xobs} \right)^2
              +\sum_p \left(B_y-B_{yobs} \right)^2
              +\sum_p \left(B_z-B_{zobs} \right)^2
        \right] \nonumber \\
L_4 &=& \left[ \sum_p \left(\Delta B_x \right)^2
                     +\left(\Delta B_y \right)^2
                     +\left(\Delta B_z \right)^2
        \right]
\end{eqnarray}
The surface integrals are here replaced by a summation $\sum_p$ over all
grid nodes $p$ of the bottom surface grid and the differentiation in the
smoothing term is achieved by the usual 5-point stencil for the 2D-Laplace
operator.
Each constraint $L_n$ is weighted by a yet undetermined factor $\mu_n$. The
first term ($n$=1) corresponds to the force-balance condition, the next
($n$=2) to the torque-free condition. The following term ($n$=3)
ensures that the optimized boundary condition agrees with the
measured photospheric data and the last terms ($n$=4) controls the
smoothing. The 2D-Laplace operator is designated by $\Delta$.

The aim of our preprocessing procedure is to minimize $L$ so that all
terms $L_n$ if possible are made small simultaneously. This will
yield a surface magnetic field
\begin{equation}
  \mathbf{B}_{\mathrm{min}} = \mathrm{argmin}(L)
\label{Bmin}
\end{equation}
Besides a dependence on the observed magnetogram, the solution
(\ref{Bmin}) now also depends on the coefficients $\mu_n$.
These coefficients are for once a formal necessity because the terms
$L_n$ physically represent different quantities with different units.
By means of these coefficients, however, we can also give more or less
weight to the individual terms in the case where a reduction in one term
contradicts the reduction in another. This competition obviously exists
between the observation term $n$=3 and the smoothing term $n$=4.

The smoothing is performed consistently for all three
magnetic field components. For this purpose we need the derivative
of $L$ with respect to each of the three field components at every
node $q$ of the bottom boundary grid.
We have, however, take account of the fact that $B_z$ is measured with
much higher accuracy than $B_x$ and $B_y$. This is achieved by
assuming the vertical component invariable compared to the horizontal
components in all terms where mixed products of the vertical and
horizontal field components occur, e.g., in the Aly constraints.
The relevant functional derivative of $L$ is therefore
\begin{eqnarray}
\frac{d L}{d(B_x)_q}
&=& 2\mu_1 \left[ \left(\sum_p B_x B_z \right) (B_z)_q
                -2\left(\sum_p B_x^2 \right) (B_x)_q
           \right] \nonumber \\
&-& 2\mu_2 \left[2\left(\sum_p x \left(B_z^2-B_x^2-B_y^2 \right) \right) (x B_x)_q
           \right. \nonumber \\ & & \left. \hspace*{1.2em}
                +2\left(\sum_p y \left(B_z^2-B_x^2-B_y^2 \right) \right) (y B_x)_q
           \right. \nonumber \\ & & \left. \hspace*{1.5em}
                 -\left(\sum_p y B_x B_z -x B_y B_z \right) (y B_z)_q
           \right] \nonumber \\
&+& 2\mu_3 \left(B_x-B_{xobs}\right)_q
\;+\;  2\mu_4  \left(\Delta\left(\Delta B_x \right)\right)_q \label{dldbx}
\end{eqnarray}

\begin{eqnarray}
\frac{d L}{d(B_y)_q}
&=& 2\mu_1 \left[ \left(\sum_p B_y B_z \right) (B_z)_q
                -2\left(\sum_p B_y^2 \right) (B_y)_q
            \right] \nonumber \\
&-& 2\mu_2 \left[2\left(\sum_p x \left(B_z^2-B_x^2-B_y^2 \right) \right) (x B_y)_q
           \right. \nonumber \\ & & \left. \hspace*{1.2em}
                +2\left(\sum_p y \left(B_z^2-B_x^2-B_y^2 \right) \right) (y B_y)_q
           \right. \nonumber \\ & & \left. \hspace*{1.5em}
                 +\left(\sum_p y B_x B_z -x B_y B_z \right) (x B_z)_q
           \right] \nonumber \\
&+& 2\mu_3 \left(B_y-B_{yobs}\right)_q
\;+\; 2\mu_4 \left(\Delta\left(\Delta B_y \right)\right)_q
\label{dldby}
\end{eqnarray}

\begin{eqnarray}
\frac{d L}{d(B_z)_q}
&=& 2\mu_3 \left(B_z-B_{zobs}\right)_q
\;+\; 2\mu_4 \left(\Delta\left(\Delta B_z \right)\right)_q
\label{dldbz}
\end{eqnarray}

The optimization is performed iteratively by a simple Newton scheme
which replaces
\begin{eqnarray}
  (B_x)_q &\leftarrow& (B_x)_q - \mu \frac{d L}{d(B_x)_q} \nonumber\\
  (B_y)_q &\leftarrow& (B_y)_q - \mu \frac{d L}{d(B_y)_q} \nonumber\\
  (B_z)_q &\leftarrow& (B_z)_q - \mu \frac{d L}{d(B_z)_q} \\
\end{eqnarray}
at every step.
The convergence of this scheme towards a solution of (\ref{Bmin})
is obvious: $L$ has to decrease monotonically at every step as long as
(\ref{dldbx}-\ref{dldbz}) has a nonzero component. These terms, however,
vanish only if an extremum of $L$ is reached.
Since $L$ is fourth order in $B$, this may not necessarily be a global
minimum, in rare cases if the step size is handled carelessly it may
even be a local maximum.
In practical calculations this should, however, not be a problem and
from our experience we rapidly obtain a minimum $\mathbf{B}_{\mathrm{min}}$
of $L$, once the parameters $\mu_n$ are specified.

What is left is a suitable recipe to choose $\mu_n$. We have four parameters
$\mu_n$ to choose and we restrict our freedom a little bit by giving the same weight
$\mu_1$ = $\mu_2 D^2$ $\equiv$ $\mu_{12}$ to the Aly momentum and torque constraints
where $D$ is the edge length of our numerical square box. Accordingly we will
combine $L_1+L_2/D^2$ $\equiv$ $L_{12}$. We are then left with three parameters only
two of which are independent since only the ratio of the parameters really counts.
Multiplying all $\mu_n$ with a common factor does not change the minimum solution
$\mathrm{B}$ of the preprocessed photospheric data. We use this fact and specify
$\mu_{12}$ to $1/B_{\mathrm{ave}}^2$ where $B_{\mathrm{ave}}$ is the average
magnetic field magnitude in our magnetogram. \footnote{Please note that
 this is equivalent to a normalization of the magnetic field with
 $B_{\mathrm{ave}}$ and the length scale with $D$. With this normalization
 we get $L_1=L_2=L_{12}=1$.}

With the parameters $\mu_3$ and $\mu_4$ we can now control the relative
influence of the observed data and the smoothing, respectively,
to each other and to the Aly constraints.
For the proper selection of these parameters we proceed in a similar
way as it is customary for regularization techniques of inversion
problems (e.g. \inlinecite{hansen:00})
For this purpose we visualize the different solutions
$\mathbf{B}_{\mathrm{min}}(\mu_3,\mu_4)$ for different $\mu$ parameters
in a phase space of $\log(L_{12})$, $\log(L_3)$ and $\log(L_4)$.
Since the $L$-terms depend analytically in $\mathbf{B}$, the solutions
$\mathbf{B}_{\mathrm{min}}$ span a continuous 2D surface in this
3D phase space.

\begin{figure}
\includegraphics[height=10cm,width=10cm]{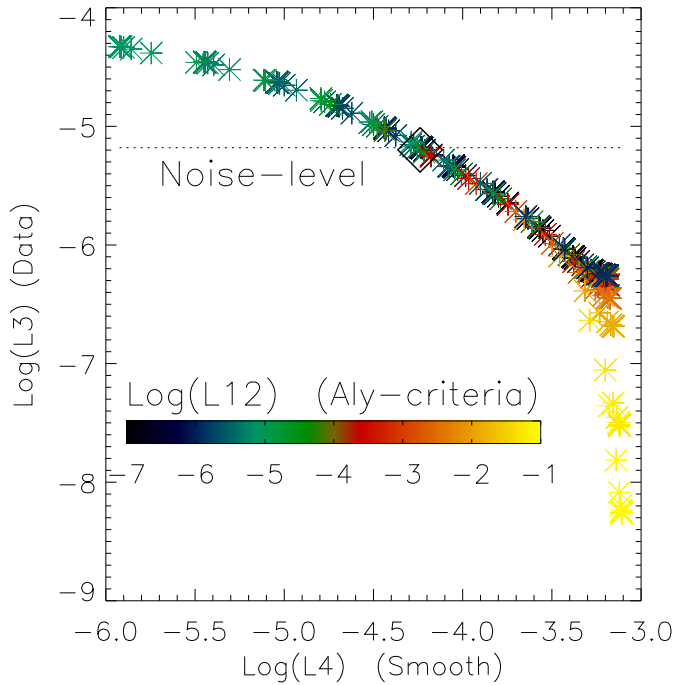}
\caption{Survey of solutions to (\ref{Bmin}) in a phase space spanned by
 $\log(L_4)$ (abscissa), $\log(L_3)$ (ordinate) and $\log(L_{12})$
 (color code). Every symbol corresponds to one solution
 $\mathbf{B}_{\mathrm{min}}$.
  The optimal parameters are marked with a rhombus. Here we have
  $L_{12}=7.6 \cdot 10^{-7}$, $L_3=6.3 \cdot 10^{-6}$ and
  $L_4=5.8 \cdot 10^{-5}$. The optimal parameters are $\mu_3=0.01$ and
  $\mu_4=0.005$.}
\label{fig0}
\end{figure}

A survey of the $\mu$ parameters for one of the example computations
(Low and Lou solution with noise model I)
described in the next section shows that
$\log(L_3(\mathbf{B}_{\mathrm{min}}))$ and
$\log(L_4(\mathbf{B}_{\mathrm{min}}))$ are almost entirely determined
by the ratio of $\mu_3$ to $\mu_4$ while
$\log(L_{12}(\mathbf{B}_{\mathrm{min}}))$ depends on the absolute
magnitude on $\mu_3$ and $\mu_4$.
It is therefore convenient to display our results in a projection
of the 3D phase space along the $\log(L_{12})$-axis as in
Fig.~\ref{fig0}. In this projection, the surface collapses nearly to a
unique curve.

An obvious limiting value is obtained for $\mu_3$ $\rightarrow$ $\infty$
which yields $\mathbf{B}_{\mathrm{min}}$ = $\mathbf{B}_{\mathrm{obs}}$
with $L_{12}$ $\sim$ 10$^{-1}$, $L_3$ = 0 and $L_4$ $\sim$ 10$^{-3}$.
For smaller values of $\mu_3$ we obtain a smoothed solution and,
depending on $\mu_4$ also satisfy the Aly criteria much better than
the original observational data.
The price we have to pay is that $L_3$ attains finite values, which
however is tolerable as long as $L_3$ does not exceed a noise value
$L_3(\mathbf{B}_{\mathrm{noise}})$ (dotted line in Fig.~\ref{fig0}).
This value in our test case is known from the amount of noise added.
For actual observations, it has to be estimated.
The intersection of our solution surface with the noise level then
defines the optimal ratio $\mu_3$ to $\mu_4$ (marked by a  rhombus
in Fig.~\ref{fig0}. We choose both numbers
small to enforce a good compliance with the Aly criteria, i.e., give
much weight to $L_{12}$. This way, $L_{12}$ is reduced by 6 orders of
magnitude, $\mathrm{B}$ is conveniently smoothed and yet it does not
deviate from the observations by more than the instrument error.
The smoothing will somewhat broaden prominent magnetic flux structures.
A careful choice of the preprocessing parameters (as described above)
ensures that the magnetic flux magnitudes and the corresponding
magnetic field topology
(which might become very complex for multiple magnetic sources,
see  \cite{schrijver:etal02}) are not affected by the preprocessing.
\section{Tests with the Low and Lou solution}
\label{sec4}
\begin{figure}
\includegraphics[bb=0 100 425 405,clip,height=6cm,width=12cm]{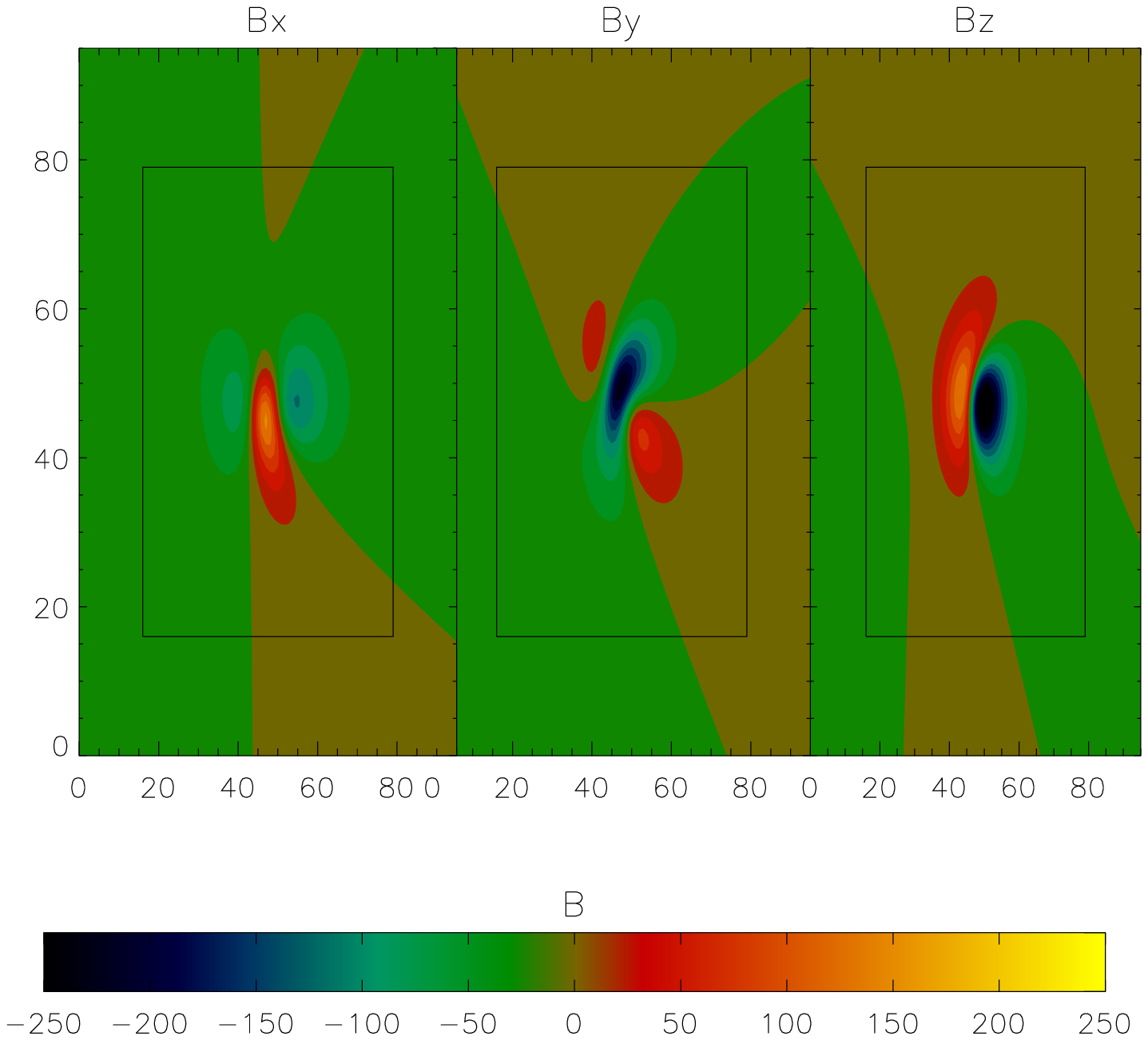}
\includegraphics[bb=0 100 425 405,clip,height=6cm,width=12cm]{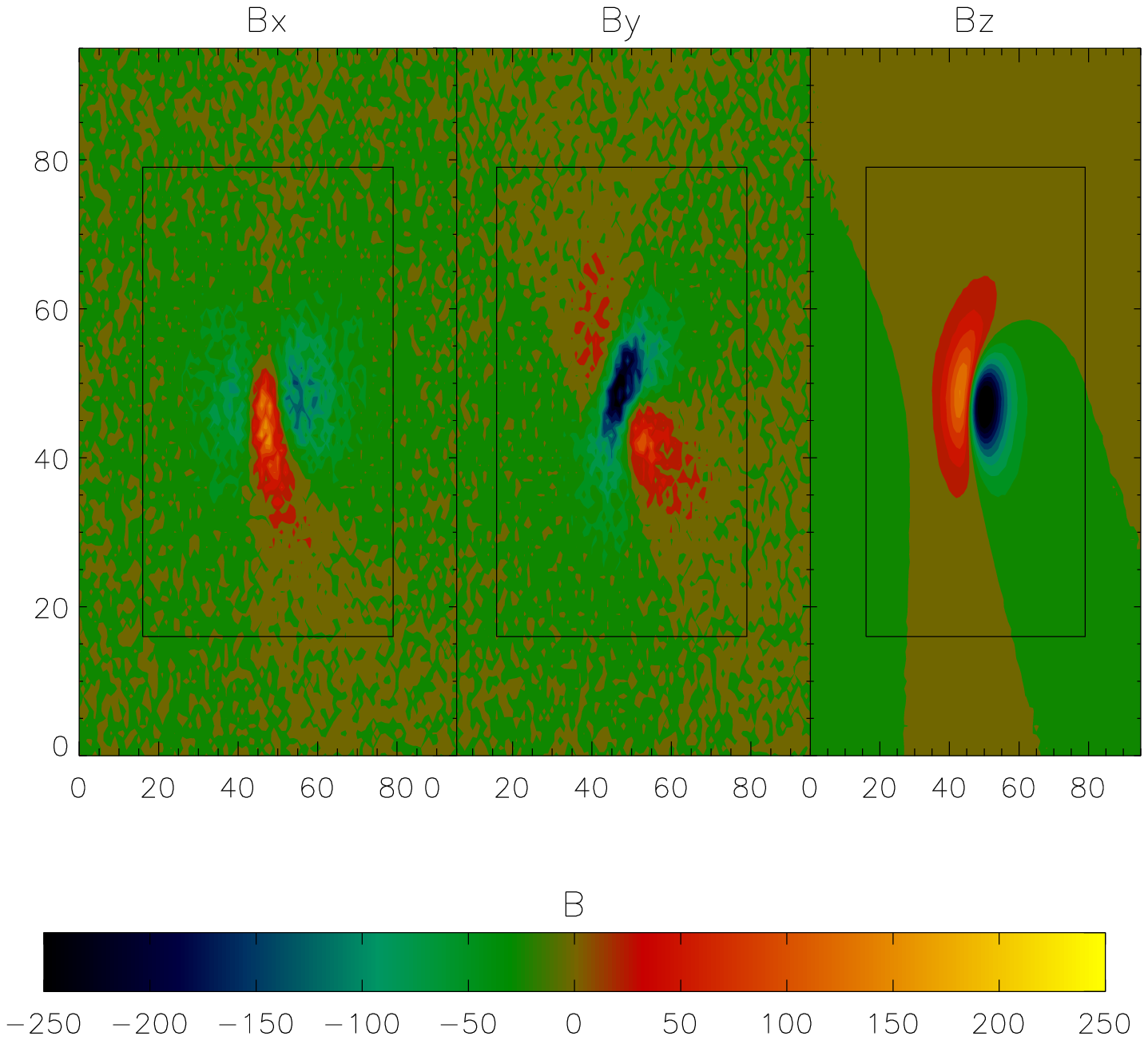}
\includegraphics[bb=0 0 425 405,clip,height=8cm,width=12cm]{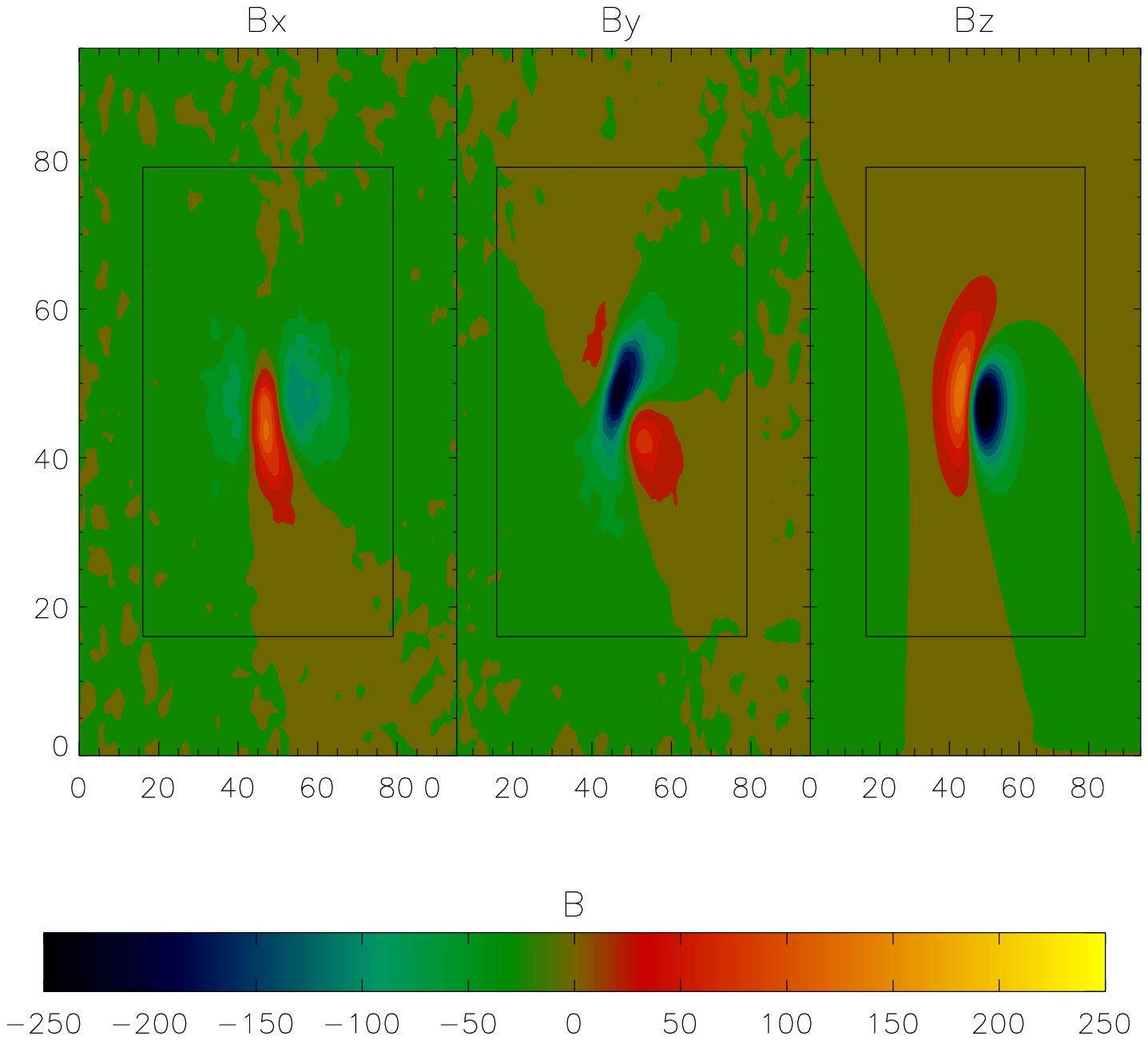}
 \caption{The first row shows the original vector magnetogram deduced from the
 Low and Lou solution. In the second row we added noise (model I) to the solution to
 simulate measurement errors. In the bottom row we applied our preprocessing
 routine to the noisy data taken from the center row. }
\label{fig1}
\end{figure}
\begin{figure}
\mbox{\includegraphics[bb=44 16 382 200,clip,height=4cm,width=7cm]{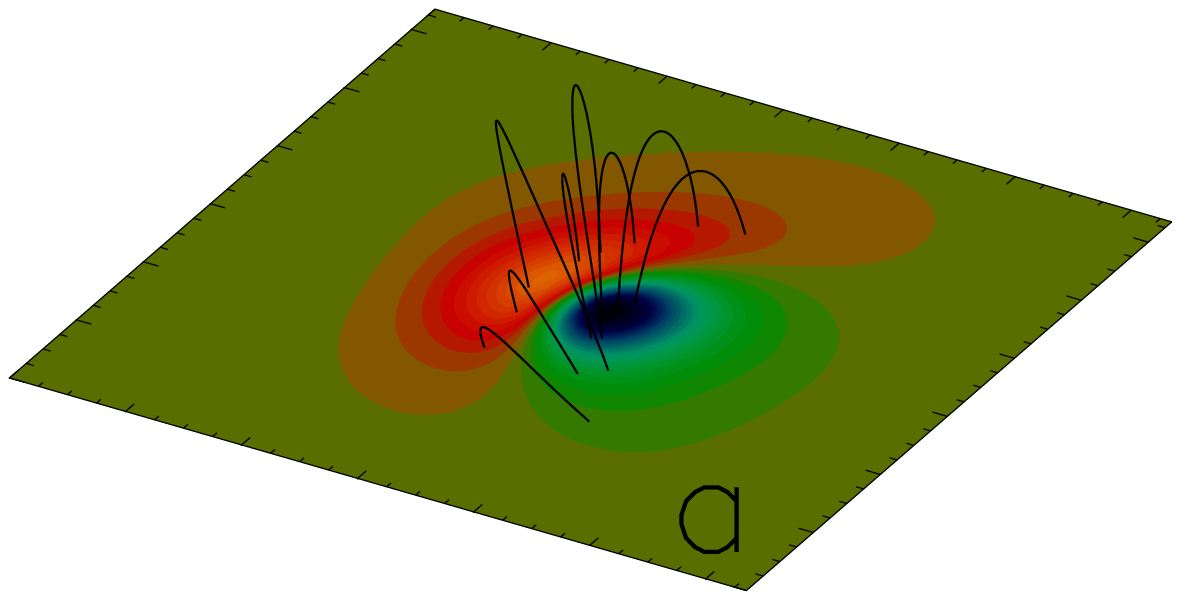}
\includegraphics[bb=44 16 382 200,clip,height=4cm,width=7cm]{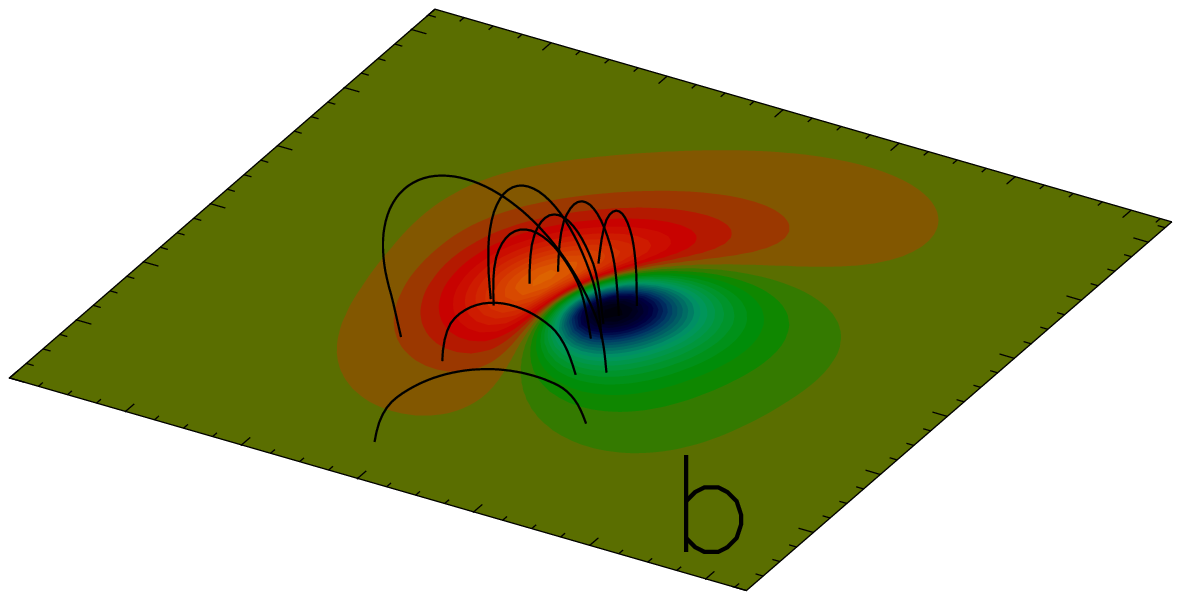}}
\mbox{\includegraphics[bb=44 16 382 200,clip,height=4cm,width=7cm]{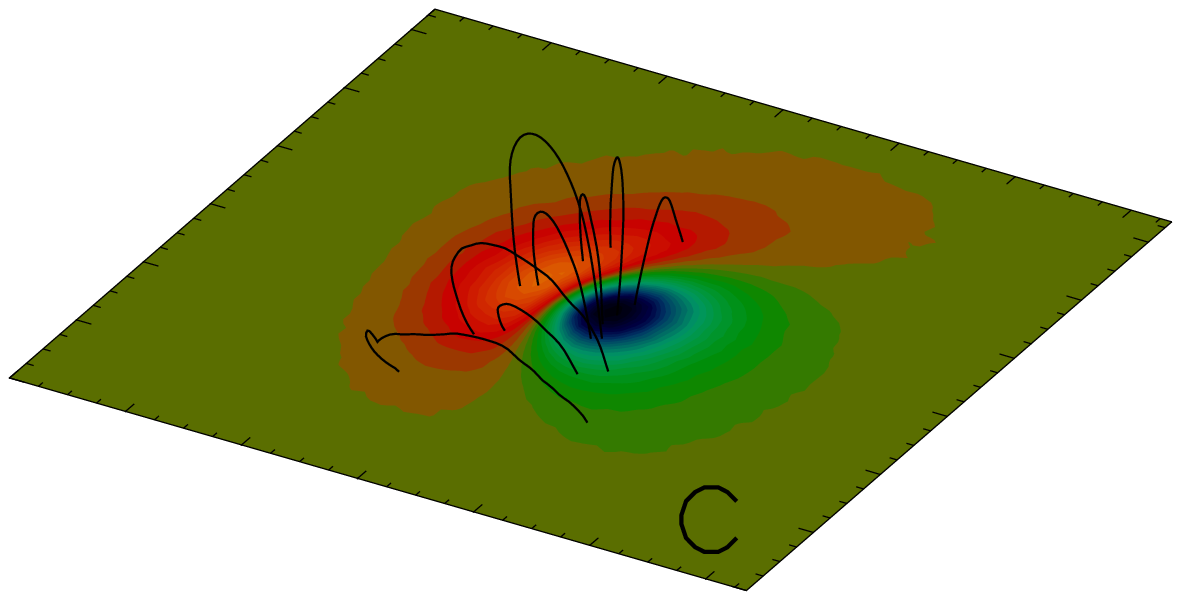}
\includegraphics[bb=44 16 382 200,clip,height=4cm,width=7cm]{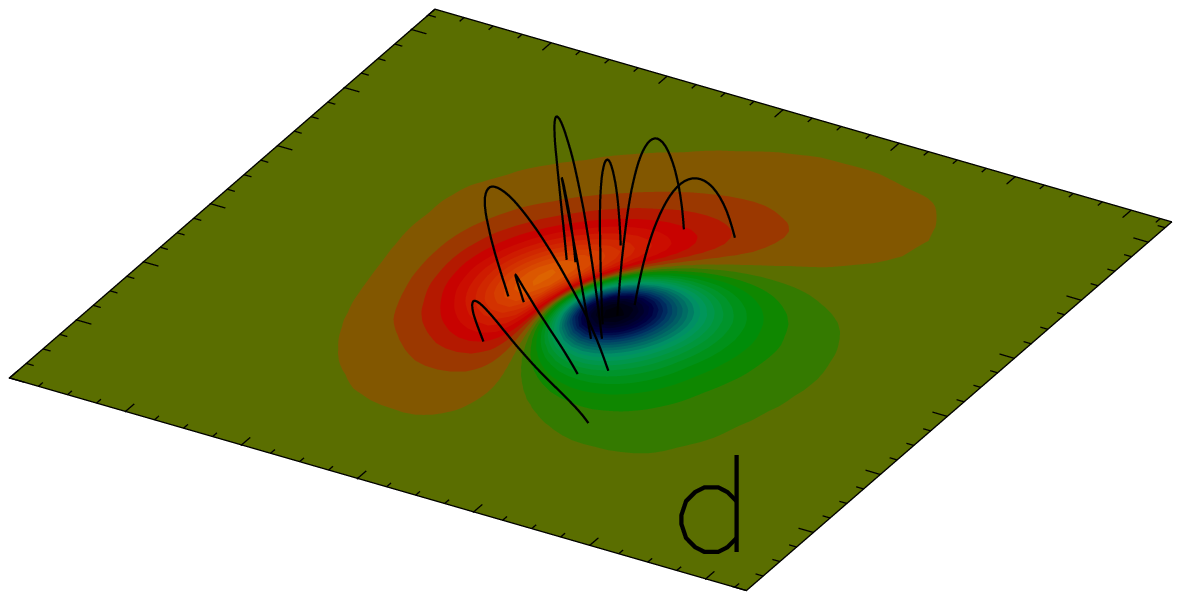}}
\mbox{\includegraphics[bb=44 16 382 200,clip,height=4cm,width=7cm]{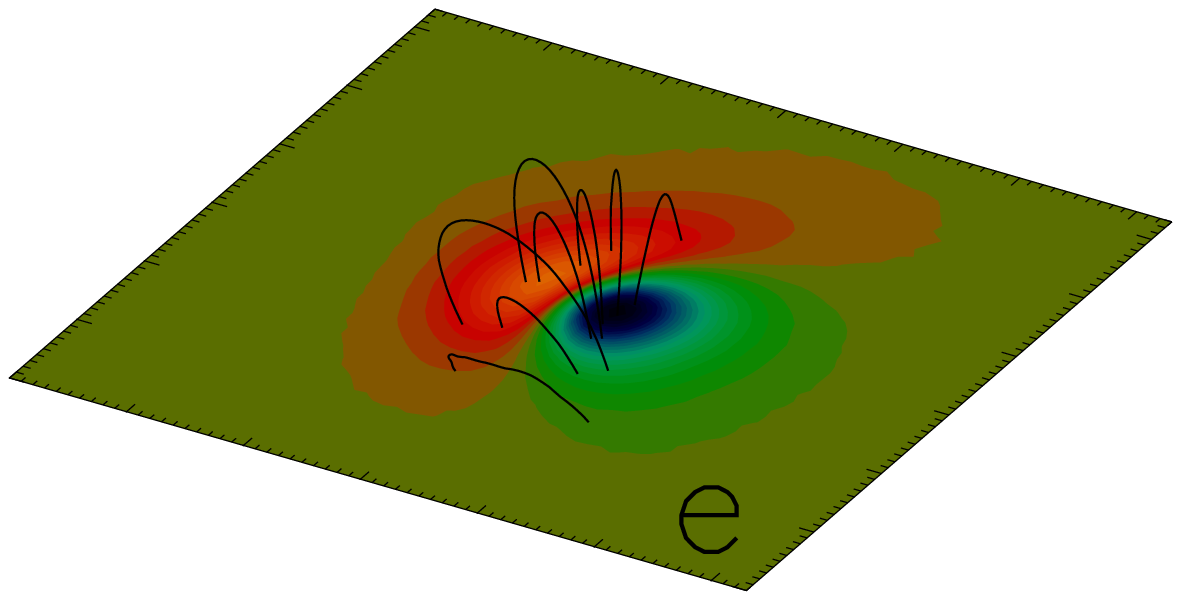}
\includegraphics[bb=44 16 382 200,clip,height=4cm,width=7cm]{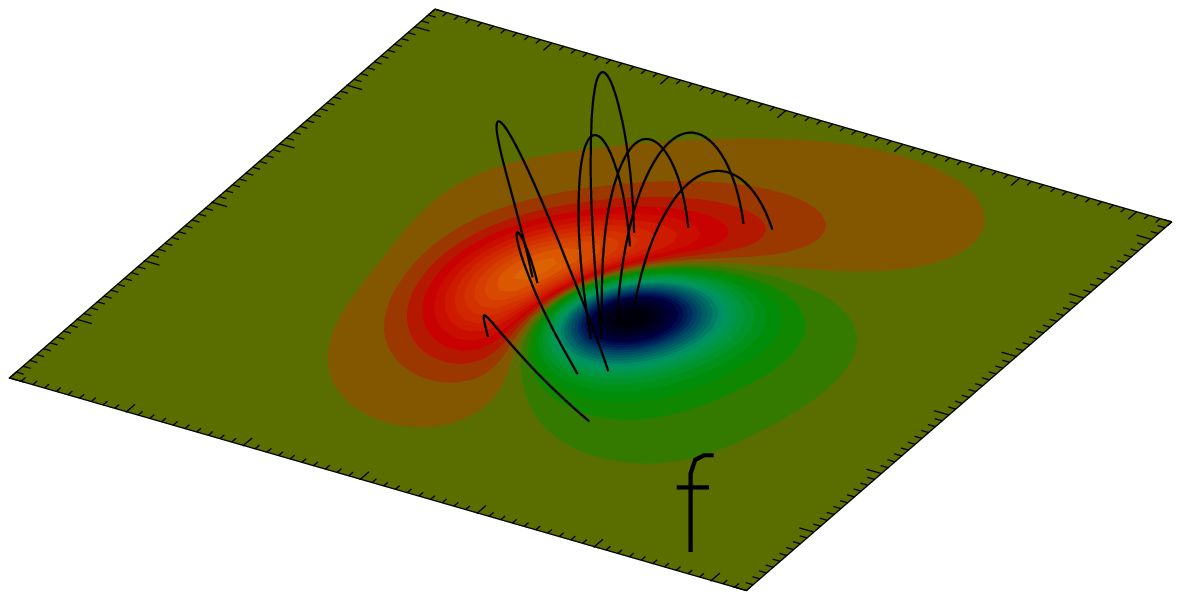}}
\mbox{\includegraphics[bb=44 16 382 200,clip,height=4cm,width=7cm]{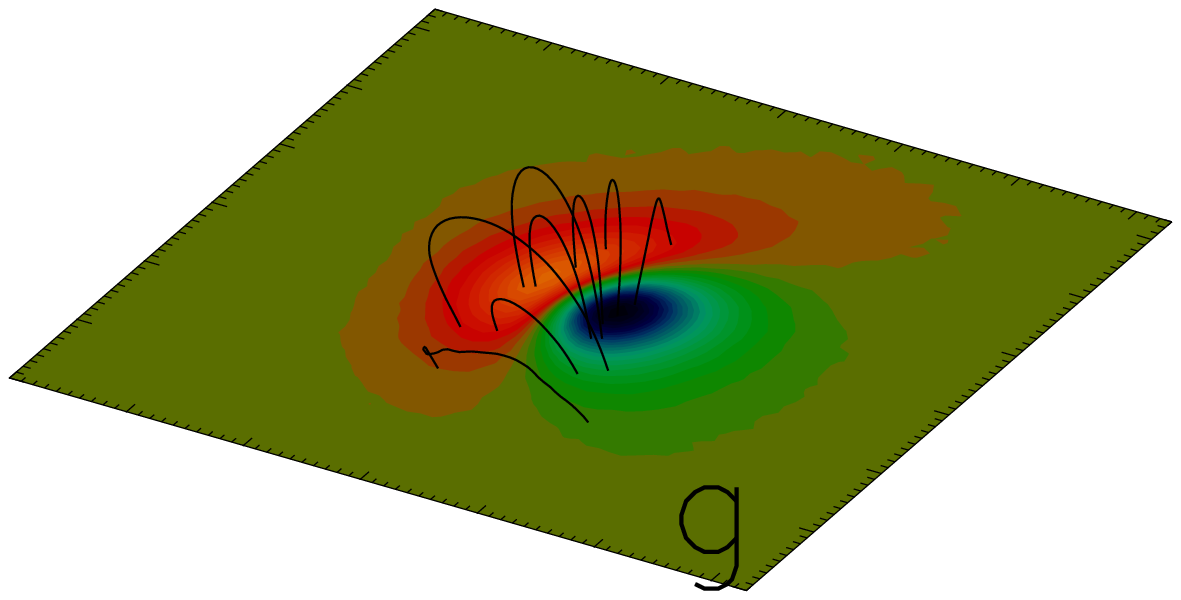}
\includegraphics[bb=44 16 382 200,clip,height=4cm,width=7cm]{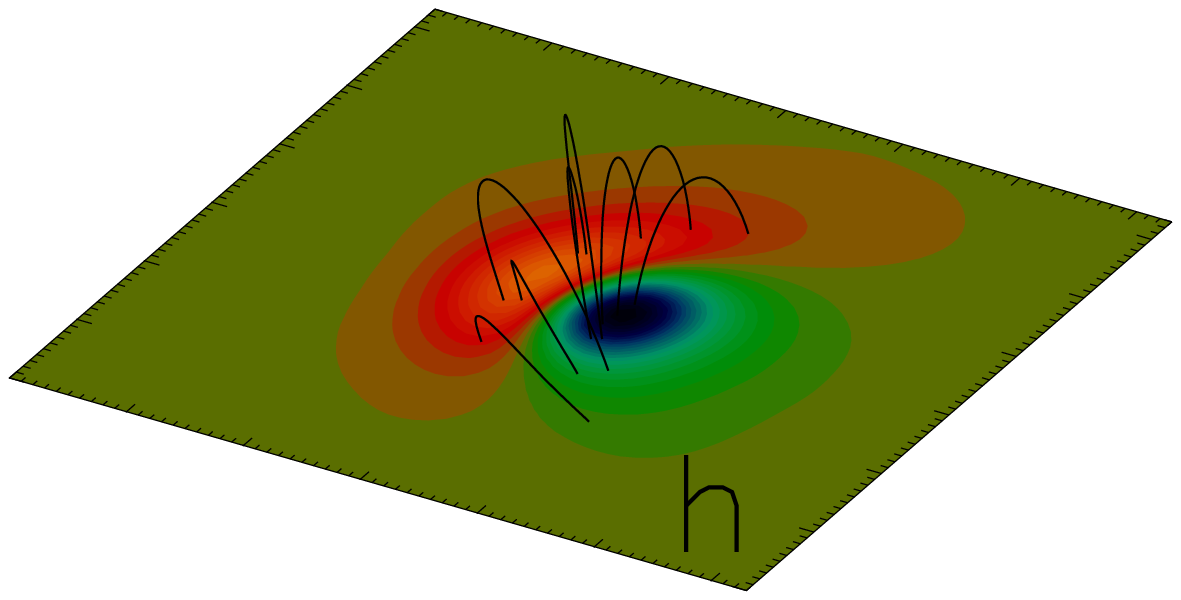}}
 \caption{ a) Some field lines for the original Low and Lou solution.
 b) Potential field reconstruction.
 c) Non-linear force-free reconstruction from noisy data (noise model I) without
 preprocessing.
 d) Non-linear force-free reconstruction from noisy data (noise model I) after
 preprocessing the vector magnetogram with our newly developed program.
 e) Non-linear force-free reconstruction from noisy data (noise model II) without
 preprocessing.
 f) Non-linear force-free reconstruction from noisy data (noise model II) after
 preprocessing.
 g) Non-linear force-free reconstruction from noisy data (noise model III) without
 preprocessing.
 h) Non-linear force-free reconstruction from noisy data (noise model III) after
 preprocessing the vector magnetogram with our newly developed program.
 The color coding shows the normal magnetic field component on the
 photosphere in all panel. One can see, that after preprocessing
 (panel d,f,h) the agreement with the original (a) is much better than
 without preprocessing (c,e,g).}
 \label{fig2}
\end{figure}
For a first test we use the semi-analytic solution found by
 \inlinecite{low:etal90}. This solution has been designed in particular
 to test non-linear force-free extrapolation codes. The solution
 contains free parameters and we use $\Phi=\pi/2$ and $l=0.3$
 (see \inlinecite{low:etal90} for details). We extract the bottom boundary
 of this equilibrium and use it as input for our extrapolation code
 (see \inlinecite{wiegelmann04}). This artificial vector magnetogram
 (see first row of Fig. \ref{fig1})
 extrapolated from a semi-analytic solution is of course in perfect
 agreement with the assumption of a force-free field above
 (Aly-criteria) and the result of our extrapolation code showed
 a reasonable agreement with the original. Real measured vector magnetograms
 are not so ideal (and smooth) of course and we simulate this effect by
 adding noise to the Low and Lou magnetogram. We add noise to this
 ideal solution in the form \\
 {\bf Noise model I:} \\
 $\delta B_i=n_l \cdot r_n \cdot \sqrt{B_i}$, where $n_l$ is the noise
 level and $r_n$ a random number in the range $-1 \dots 1$. The noiselevel
 was choosen $n_l=5.0$ for the transversal magnetic field ($B_x$, $B_y$)
 and $n_l=0.25$ for $B_z$.
 This mimics a real magnetogram (see center row of Fig. \ref{fig1}) which has
 significantly higher noise in the transversal components of the
 magnetic field. \\
 {\bf Noise model II:} \\
 $\delta B_i=n_l \cdot r_n $, where $n_l$ is the noise
 level and $r_n$ a random number in the range $-1 \dots 1$. The noiselevel
 was choosen $n_l=10.0$ for the transversal magnetic field ($B_x$, $B_y$)
 and $n_l=0.5$ for $B_z$. This noise model adds an additive noise, independent
 from the magnetic field strength. \\
 {\bf Noise model III:} \\
 $\delta B_z={\rm constant}$,
${\delta B_t}=\frac{\delta B_{{\rm tmin}}^2}{\sqrt{B_t^2+\delta B_{{\rm
tmin}}^2}}$
where we choose the constant noise level of $B_z$ to $1$ and the minimum
detection level $\delta B_{{\rm tmin}}=20$. This noise model mimics the
effect that the transversal noise level is higher in regions with a low magnetic
field strength, which is an immediate consequence of Eq. \ref{eq4}. \\

 The bottom
 row of Fig. \ref{fig1} shows the preprocessed vector magnetogram
 (for noise model I) after
 applying our procedure. The aim of the preprocessing is to use the
 resulting magnetogram as input for a non-linear force-free magnetic
 field extrapolation. Fig. \ref{fig2} shows in panel a) the
 original Low and Lou solution and in panel b) a corresponding potential
 field reconstruction. In Fig. \ref{fig2} we present only the
 central region of the whole magnetogram (marked with black rectangular
 box in Fig. \ref{fig1} because the surrounding magnetogram is used
 as a boundary layer ($16$ grid points) for our non-linear force-free code
 (Our non-linear force-free code is explained in detail in \inlinecite{wiegelmann04}).
 The computation was done on a $96 \times 96 \times 80$ grid including a $16$ pixel
 boundary layer towards the lateral and top boundary of the computational box.)
 In the remaining panels of Fig. \ref{fig2} we demonstrate the effect of
 the different noise models on the reconstruction. The noise levels were
 chosen so that the mean noise was similar for all three noise models.
 Fig. \ref{fig2} c) shows a non-linear force-free reconstruction with noisy
 data (noise model I, magnetogram shown in the center panel of Fig. \ref{fig1}) and Fig. \ref{fig2} d)
 presents a non-linear force-free reconstruction after preprocessing
 (Magnetogram shown in the bottom panel of Fig. \ref{fig1}). After
 preprocessing(Fig. \ref{fig2} d) we get a much better agreement with the original
 solution (Fig. \ref{fig2} a). Fig. \ref{fig2} e) and f) show a non-linear
 force-free reconstruction for noise model II without (e) and after (f)
 preprocessing of the noisy data.
 Fig. \ref{fig2} g) and h) show a non-linear
 force-free reconstruction for noise model III without (g) and after (h)
 preprocessing of the noisy data.
 Similar as for noise model I we find that
 the preprocessed data agree  better with the original Fig. \ref{fig2} a).
 We check the correlation of the original solution with our reconstruction with
 help of the vector correlation function $VC$  defined as
\begin{equation}
VC=\frac{\sum_i {\bf v_i} \cdot {\bf w_i}}{\sqrt{\sum_i |{\bf v_i}|^2} \; \sqrt{\sum_i |{\bf w_i}|^2}}.
\end{equation}
${\bf v_i}$ correspond to a reference field (Low and Lou solution) and ${\bf w_i}$ to
the non-linear force-free field reconstructed with our code.

\begin{tabular}{cccc}
Case & remark & preprocessing & vector correlation \\
 \hline
 a) & reference & & 1.00 \\
 b) & potential & & 0.85 \\
 c) & Noise model I & No & 0.94 \\
 d) & Noise model I & Yes &0.98 \\
 e) & Noise model II & No &0.95 \\
 f) & Noise model II & Yes &0.98 \\
 g) & Noise model III & No &0.93 \\
 h) & Noise model III & Yes &0.98 \\
\end{tabular}
$ $ \\
 The table confirms the visual inspection of Fig. \ref{fig2}. The correlation of
 the reconstructed magnetic field with the original becomes significantly improved
 after preprocessing of the data for all noise models.
 We knew already from previous studies
 \cite{wiegelmann:etal03,wiegelmann04} that noise and inconsistencies
 in vector magnetograms have negative influence on the non-linear force-free
 reconstruction and the preprocessing routine described in this work
 tells us how to overcome these difficulties.

\section{Application to data from the Solar Flare Telescope (SFT)}
\label{sec5}
\begin{figure}
\includegraphics[bb=0 100 425 405,clip,height=6cm,width=12cm]{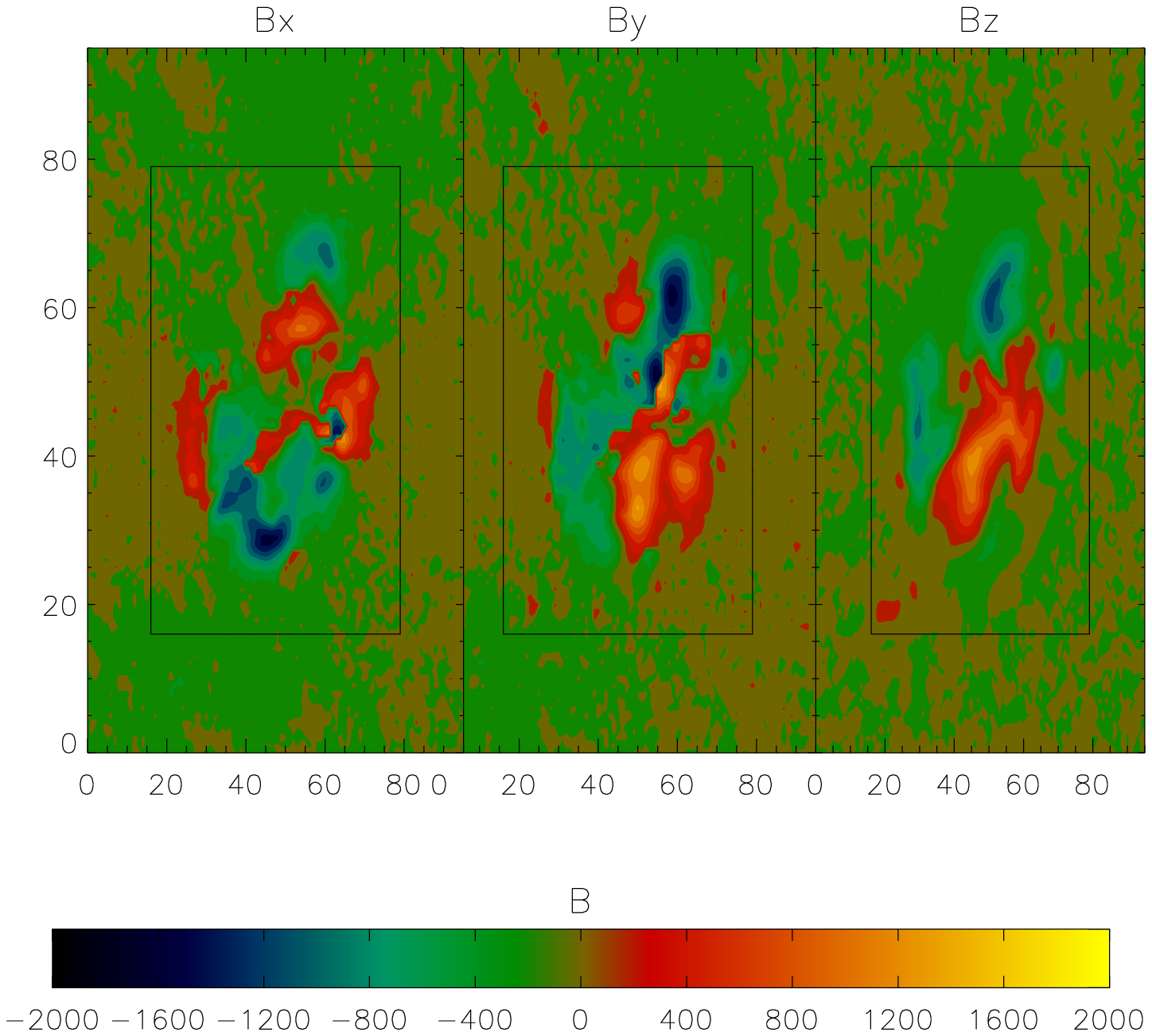}
\includegraphics[bb=0 0 425 405,clip,height=8cm,width=12cm]{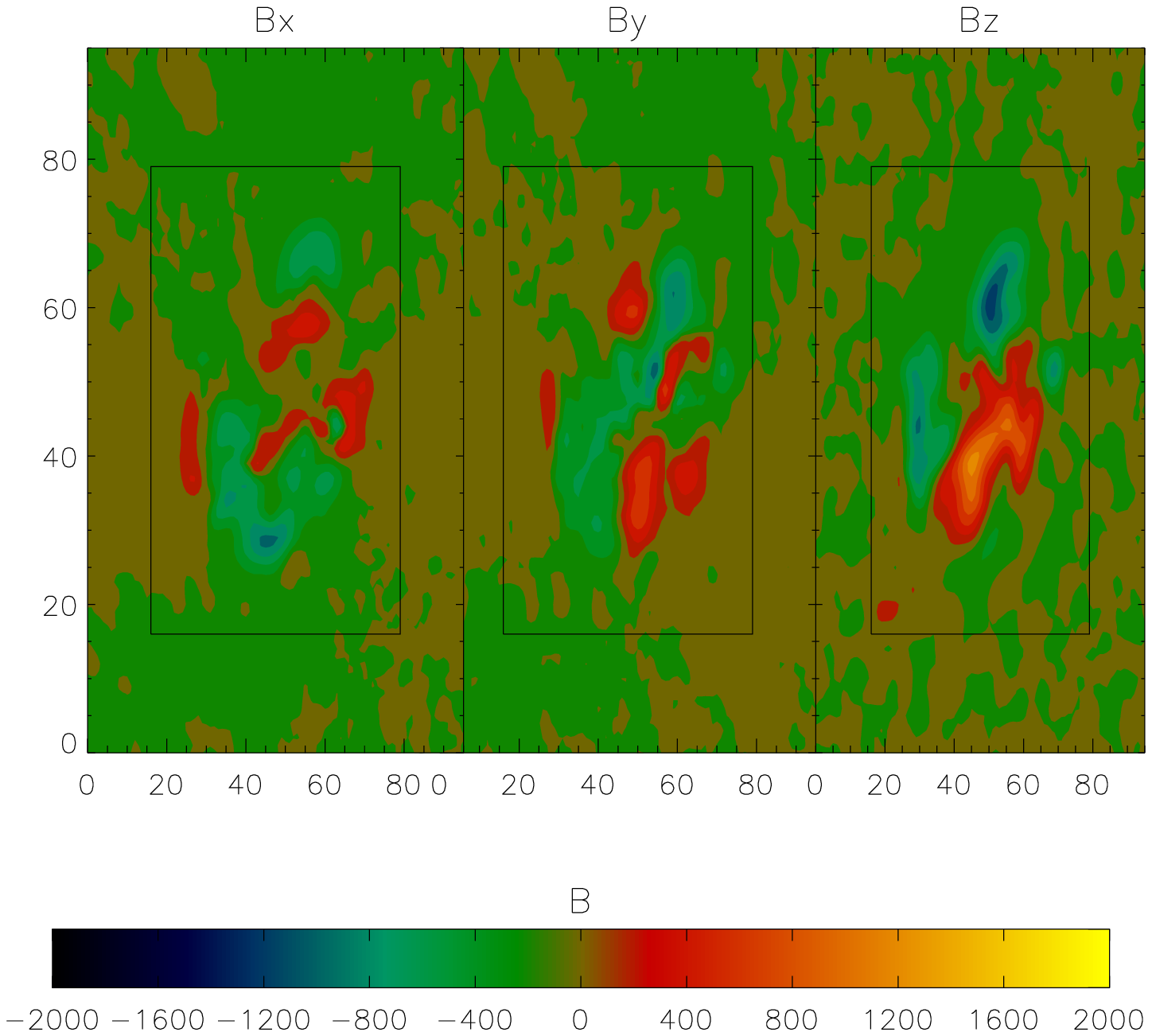}
 \caption{Top panel: vector magnetogram of AR 7321 taken with SFT at october, 26 1992.
 Bottom panel: After applying our preprocessing procedure.
 The resolution of the original SFT-magnetogram has been reduced by
 a factor of two to speed up the non-linear force-free magnetic field
 reconstruction.
 An inspection of the top and bottom panel shows that local small scale structures
with strong flux density in the original lead to a broader structure with
weaker flux density after preprocessing. Sharp boundaries between positive and
negative flux regions are also broadened due to smoothing during preprocessing.}
\label{fig3}
\end{figure}
\begin{figure}
\mbox{\includegraphics[bb=44 16 382 250,clip,height=5cm,width=7cm]{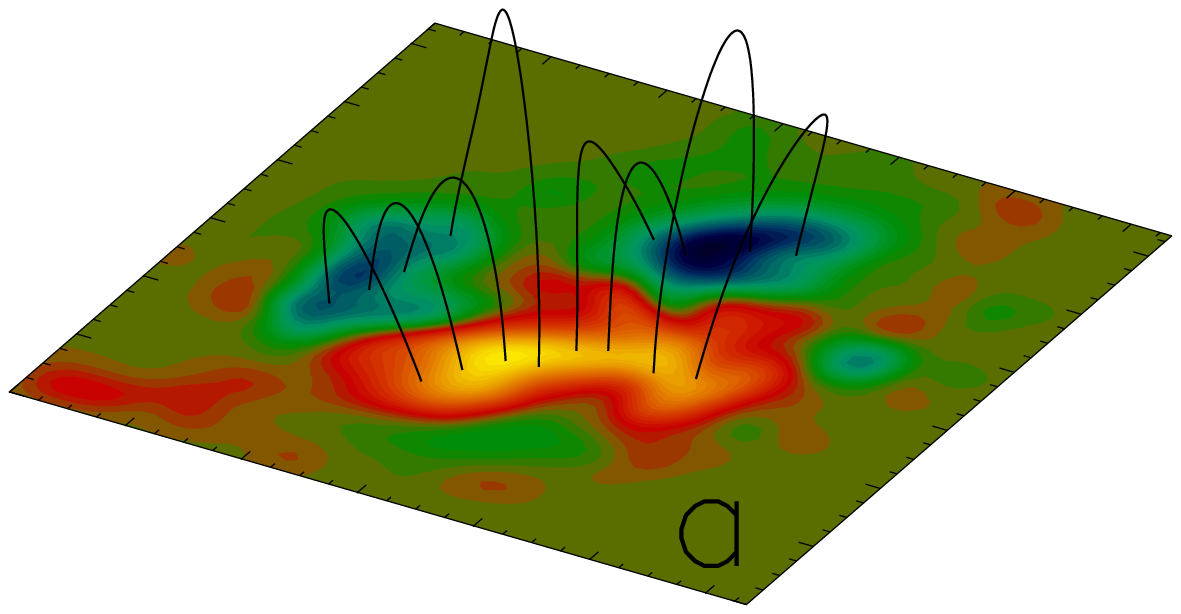}
\includegraphics[bb=44 16 382 250,clip,height=5cm,width=7cm]{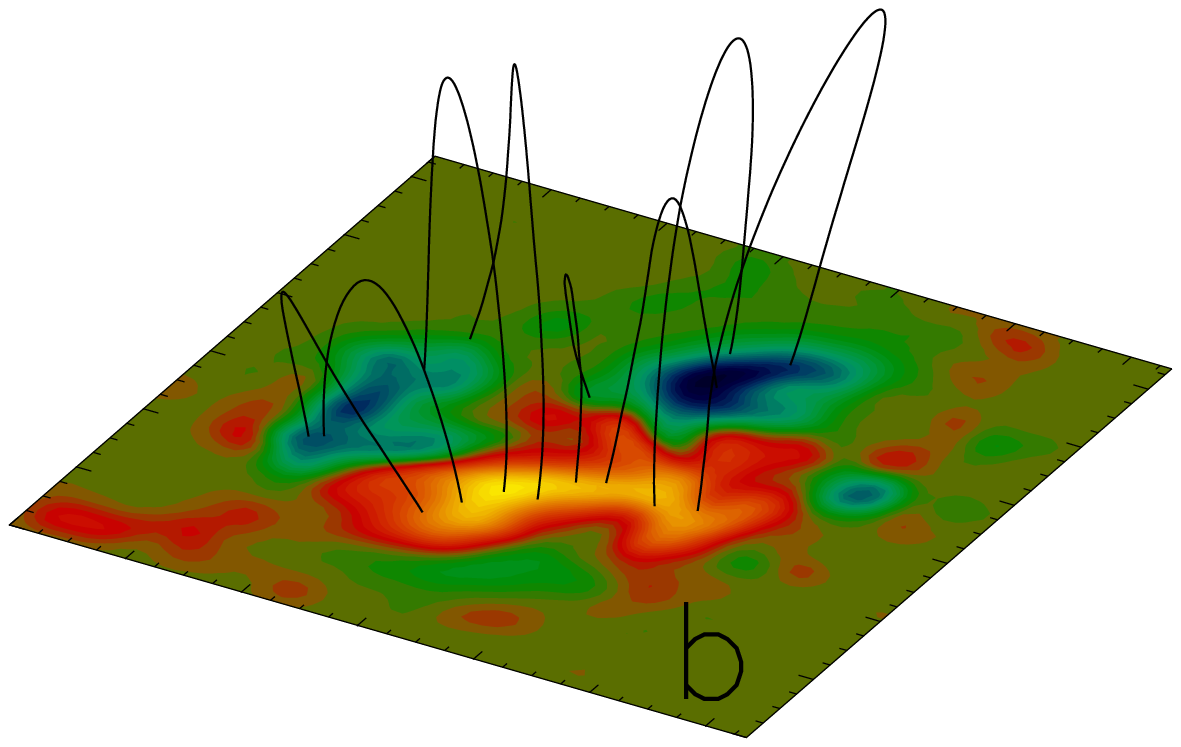}}
\caption{AR 7321 measured with SFT-data from october, 26 1992.
a) Potential field reconstruction.
b)Non-linear force-free reconstruction. The field lines start from the same
footpoints within regions $B_z>0$ in both panels.}
\label{fig4}
\end{figure}
 In this section we apply our method to vector magnetogram data
 of AR 7321 taken with
 SFT at the National Astronomical Observatory (NAO) in Tokyo.
 The SFT instrument is described in
 \cite{sakurai:etal95} and the evolution of the photospheric
 magnetic field in AR 7321 has been studied by \inlinecite{li:etal00b}.
 The top panel of Fig. \ref{fig3}  shows the original data
 (original resolution reduced by a factor of two.) and the bottom panel after
 preprocessing. The structure of all components of the magnetic field is similar
 to the original. The magnetogram is almost flux-balanced
 ($\epsilon_{\mbox{flux}}=0.045$), but
 Aly's criteria are not fulfilled in the original
 ($\epsilon_{\mbox{Aly}}=1.50$). After preprocessing we get
 $\epsilon_{\mbox{Aly}}=0.029$
 and the data are also somewhat smoother than the original.
 The L-values of the original data and after preprocessing are:
 \begin{tabular}{llll}
 &&original & preprocessed \\
 $L_{12}$&(Aly-criteria) & $0.58$ & $3.6 \cdot 10^{-5}$ \\
 $L_3$&      (Data)      & $0.0$ & $1.0 \cdot 10^{-5}$ \\
 $L_4$&    (Smoothing)   & $2.1 \cdot 10^{-4}$ & $5.2 \cdot 10^{-6}$
 \end{tabular}

  Fig. \ref{fig4} a)
 shows a potential field reconstruction and Fig. \ref{fig4} b) a non-linear
 force-free reconstruction based on the preprocessed vector magnetogram.
 The non-linear force-free computation was done in a $96 \times 96 \times 80$
 box, including a boundary layer of $16$ grid points towards the lateral and
 top boundary. Fig. \ref{fig4} shows only the center cube of
 $64 \times 64 \times 64$ pixel which corresponds to the area marked with
 a black rectangular box in Fig. \ref{fig3}.
\clearpage
\section{Conclusions}
\label{sec6}
 Within this work we presented a method for the preprocessing of
 vector magnetogram data with the aim to use the result of the
 preprocessing as input for a non-linear force-free magnetic
 field extrapolation with help of an optimization code method.
 As a first test of the method we use the Low and Lou solution
 with overlaid noise with different noise models. A direct use
 of the noisy photospheric data for a non-linear force-free
 extrapolation showed no good agreement with the original
 Low and Lou solution, but after applying our newly developed
 preprocessing method we got a reasonable agreement with the original.
 The preprocessing method uses the high noise level in the transversal
 vector magnetogram components to drive the magnetogram towards
 boundary conditions which are consistent with the assumption of
 a force-free field above. To do so we use an minimization principle.
 On the one hand we control that the final boundary data are as close
 as possible (within the noise level) to the original measured data
 and on the other hand the data are forced to fulfill the Aly-criteria
 and be sufficiently smooth. Smoothness of the boundary data is
 required for the non-linear force-free extrapolation code, but also
 physically motivated because the magnetic field at the basis of
 the corona should be smoother than on the photosphere, where it
 is measured.

 Let us remark that non-linear force-free extrapolations are only valid as
 long as the plasma $\beta$ is low and non-magnetic forces like pressure gradients
 and gravity can be neglected.
This is usually the case in the corona (low $\beta$ plasma, $\beta\approx 10^{-4}$),
but not in the photosphere ($\beta \approx 1$).
A fully understanding on how the magnetic field evolves from the (non force-free)
photosphere through the chromosphere and transition region into the (force-free)
corona would require more advanced models, at least magnetohydrostatic equilibria.
Measurements of the photospheric magnetic field vector are not sufficient for such
models and one needs additional observations regarding plasma quantities
(e.g. density, temperature, pressure).
A magnetohydrostatic approach is also
required to model prominences where the magnetic field
counteracts the gravitational force of the material they support.

 We apply our newly
 developed preprocessing program to data taken with the Solar Flare
 Telescope and use the preprocessed boundary data for a non-linear
 force-free field extrapolation.
\begin{acknowledgements}
The work of T. Wiegelmann was supported by  DLR-grant 50 OC 0007 and
a JSPS visitor grant at the National Astronomical Observatory in Tokyo.
We thank the referee Bruce Lites for useful remarks.
\end{acknowledgements}

\begin{thebibliography}{}

\bibitem[\protect\citeauthoryear{{Alissandrakis}}{1981}]{alissandrakis81}
{Alissandrakis}, C.~E.: 1981, `{On the computation of constant alpha force-free
  magnetic field}'.
\newblock {\em A\&A} {\bf 100}, 197--200.

\bibitem[\protect\citeauthoryear{{Aly}}{1989}]{aly89}
{Aly}, J.~J.: 1989, `{On the reconstruction of the nonlinear force-free coronal
  magnetic field from boundary data}'.
\newblock {\em Sol. Phys.} {\bf 120}, 19--48.

\bibitem[\protect\citeauthoryear{{Amari} et~al.}{1997}]{amari:etal97}
{Amari}, T., J.~J. {Aly}, J.~F. {Luciani}, T.~Z. {Boulmezaoud}, and Z. {Mikic}:
  1997, `{Reconstructing the Solar Coronal Magnetic Field as a Force-Free
  Magnetic Field}'.
\newblock {\em Sol. Phys.} {\bf 174}, 129--149.

\bibitem[\protect\citeauthoryear{{Arnaud} and {Newkirk}}{1987}]{arnaud:etal87}
{Arnaud}, J. and G. {Newkirk}: 1987, `{Mean properties of the polarization of
  the Fe XIII 10747 A coronal emission line}'.
\newblock {\em A\&A} {\bf 178}, 263--268.

\bibitem[\protect\citeauthoryear{{Chiu} and {Hilton}}{1977}]{chiu:etal77}
{Chiu}, Y.~T. and H.~H. {Hilton}: 1977, `{Exact Green's function method of
  solar force-free magnetic-field computations with constant alpha. I - Theory
  and basic test cases}'.
\newblock {\em ApJ} {\bf 212}, 873--885.

\bibitem[\protect\citeauthoryear{{Cuperman} et~al.}{1989}]{cuperman:etal89}
{Cuperman}, S., L. {Ofman}, and M. {Semel}: 1989, `{Determination of
  constant-alpha force-free magnetic fields above the photosphere using
  three-component boundary conditions}'.
\newblock {\em A\&A} {\bf 216}, 265--277.

\bibitem[\protect\citeauthoryear{{Demoulin} and
  {Priest}}{1992}]{demoulin:etal92}
{Demoulin}, P. and E.~R. {Priest}: 1992, `{The properties of sources and sinks
  of a linear force-free field}'.
\newblock {\em A\&A} {\bf 258}, 535--541.

\bibitem[\protect\citeauthoryear{{Gary}}{2001}]{gary01}
{Gary}, G.~A.: 2001, `{Plasma Beta above a Solar Active Region: Rethinking the
  Paradigm}'.
\newblock {\em Sol. Phys.} {\bf 203}, 71--86.

\bibitem[\protect\citeauthoryear{Hansen}{2000}]{hansen:00}
Hansen, P.~C.: 2000, `The {L-}curve and its use in the numerical treatment of
  inverse problems'.
\newblock In: P. Johnston (ed.): {\em InviteComputational Inverse Problems in
  Electrocardiology}. {WIT} Press.

\bibitem[\protect\citeauthoryear{{House}}{1977}]{house77}
{House}, L.~L.: 1977, `{Coronal emission-line polarization from the statistical
  equilibrium of magnetic sublevels. I - Fe XIII}'.
\newblock {\em ApJ} {\bf 214}, 632--652.

\bibitem[\protect\citeauthoryear{{Judge}}{1998}]{judge98}
{Judge}, P.~G.: 1998, `{Spectral Lines for Polarization Measurements of the
  Coronal Magnetic Field. I. Theoretical Intensities}'.
\newblock {\em ApJ} {\bf 500}, 1009--+.

\bibitem[\protect\citeauthoryear{{Lagg} et~al.}{2004}]{lagg:etal04}
{Lagg}, A., J. {Woch}, N. {Krupp}, and S.~K. {Solanki}: 2004, `{Retrieval of
  the full magnetic vector with the He I multiplet at 1083 nm. Maps of an
  emerging flux region}'.
\newblock {\em A\&A} {\bf 414}, 1109--1120.

\bibitem[\protect\citeauthoryear{{Li} et~al.}{2000}]{li:etal00b}
{Li}, H., T. {Sakurai}, K. {Ichimoto}, and S. {UeNo}: 2000, `{Magnetic Field
  Evolution Leading to Solar Flares II. Cases with High Magnetic Shear and
  Flare-Related Shear Change}'.
\newblock {\em Publ. Astron. Soc. Japan} {\bf 52}, 483--497.

\bibitem[\protect\citeauthoryear{{Lin} et~al.}{2004}]{lin:etal04}
{Lin}, H., J.~R. {Kuhn}, and R. {Coulter}: 2004, `{Coronal Magnetic Field
  Measurements}'.
\newblock {\em ApJ} {\bf 613}, L177--L180.

\bibitem[\protect\citeauthoryear{{Low} and {Lou}}{1990}]{low:etal90}
{Low}, B.~C. and Y.~Q. {Lou}: 1990, `{Modeling solar force-free magnetic
  fields}'.
\newblock {\em ApJ} {\bf 352}, 343--352.

\bibitem[\protect\citeauthoryear{{Metcalf}}{1994}]{metcalf94}
{Metcalf}, T.~R.: 1994, `{Resolving the 180-degree ambiguity in vector magnetic
  field measurements: The 'minimum' energy solution}'.
\newblock {\em Sol. Phys.} {\bf 155}, 235--242.

\bibitem[\protect\citeauthoryear{{R{\' e}gnier} et~al.}{2002}]{regnier:etal02}
{R{\' e}gnier}, S., T. {Amari}, and E. {Kersal{\' e}}: 2002, `{3D Coronal
  magnetic field from vector magnetograms: non-constant-alpha force-free
  configuration of the active region NOAA 8151}'.
\newblock {\em A\&A} {\bf 392}, 1119--1127.

\bibitem[\protect\citeauthoryear{{Roumeliotis}}{1996}]{roumeliotis96}
{Roumeliotis}, G.: 1996, `{The ``Stress-and-Relax'' Method for Reconstructing
  the Coronal Magnetic Field from Vector Magnetograph Data}'.
\newblock {\em ApJ} {\bf 473}, 1095--+.

\bibitem[\protect\citeauthoryear{{Sakurai}}{1981}]{sakurai81}
{Sakurai}, T.: 1981, `{Calculation of Force-Free Magnetic Field with Non
  Constant Alpha}'.
\newblock {\em Sol. Phys.} {\bf 69}, 343--+.

\bibitem[\protect\citeauthoryear{{Sakurai} et~al.}{1995}]{sakurai:etal95}
{Sakurai}, T., K. {Ichimoto}, Y. {Nishino}, K. {Shinoda}, M. {Noguchi}, E.
  {Hiei}, T. {Li}, F. {He}, W. {Mao}, H. {Lu}, G. {Ai}, Z. {Zhao}, S.
  {Kawakami}, and J. {Chae}: 1995, `{Solar flare telescope at Mitaka}'.
\newblock {\em Publ. Astron. Soc. Japan} {\bf 47}, 81--92.

\bibitem[\protect\citeauthoryear{{Schmidt}}{1964}]{schmidt64}
{Schmidt}, H.~U.: 1964, `{On the Observable Effects of Magnetic Energy Storage
  and Release Connected With Solar Flares}'.
\newblock In: {\em The Physics of Solar Flares}. pp. 107--+.

\bibitem[\protect\citeauthoryear{{Schrijver} and
  {Title}}{2002}]{schrijver:etal02}
{Schrijver}, C.~J. and A.~M. {Title}: 2002, `{The topology of a mixed-polarity
  potential field, and inferences for the heating of the quiet solar corona}'.
\newblock {\em Sol. Phys.} {\bf 207}, 223--240.

\bibitem[\protect\citeauthoryear{{Seehafer}}{1978}]{seehafer78}
{Seehafer}, N.: 1978, `{Determination of constant alpha force-free solar
  magnetic fields from magnetograph data}'.
\newblock {\em Sol. Phys.} {\bf 58}, 215--223.

\bibitem[\protect\citeauthoryear{{Seehafer}}{1982}]{seehafer82}
{Seehafer}, N.: 1982, `{A comparison of different solar magnetic field
  extrapolation procedures}'.
\newblock {\em Sol. Phys.} {\bf 81}, 69--80.

\bibitem[\protect\citeauthoryear{{Semel}}{1967}]{semel67}
{Semel}, M.: 1967, `{Contribution {\` a} l{\' e}tude des champs magn{\'
  e}tiques dans les r{\' e}gions actives solaires}'.
\newblock {\em Annales d'Astrophysique} {\bf 30}, 513--513.

\bibitem[\protect\citeauthoryear{{Semel}}{1988}]{semel88}
{Semel}, M.: 1988, `{Extrapolation functions for constant-alpha force-free
  fields - Green's method for the oblique boundary value}'.
\newblock {\em A\&A} {\bf 198}, 293--299.

\bibitem[\protect\citeauthoryear{{Solanki} et~al.}{2003}]{solanki:etal03}
{Solanki}, S.~K., A. {Lagg}, J. {Woch}, N. {Krupp}, and M. {Collados}: 2003,
  `{Three-dimensional magnetic field topology in a region of solar coronal
  heating}'.
\newblock {\em Nature} {\bf 425}, 692--695.

\bibitem[\protect\citeauthoryear{{Valori} et~al.}{2005}]{valori:etal05}
{Valori}, G., B. {Kliem}, and R. {Keppens}: 2005, `{Extrapolation of a
  nonlinear force-free field containing a highly twisted magnetic loop}'.
\newblock {\em A\&A} {\bf 433}, 335--347.

\bibitem[\protect\citeauthoryear{{Wheatland}}{2004}]{wheatland04}
{Wheatland}, M.~S.: 2004, `{Parallel Construction of Nonlinear Force-Free
  Fields}'.
\newblock {\em Sol. Phys.} {\bf 222}, 247--264.

\bibitem[\protect\citeauthoryear{{Wheatland} et~al.}{2000}]{wheatland:etal00}
{Wheatland}, M.~S., P.~A. {Sturrock}, and G. {Roumeliotis}: 2000, `{An
  Optimization Approach to Reconstructing Force-free Fields}'.
\newblock {\em ApJ} {\bf 540}, 1150--1155.

\bibitem[\protect\citeauthoryear{{Wiegelmann}}{2004}]{wiegelmann04}
{Wiegelmann}, T.: 2004, `{Optimization code with weighting function for the
  reconstruction of coronal magnetic fields}'.
\newblock {\em Sol. Phys.} {\bf 219}, 87--108.

\bibitem[\protect\citeauthoryear{{Wiegelmann} et~al.}{2005}]{wiegelmann:etal05}
{Wiegelmann}, T., A. {Lagg}, S.~K. {Solanki}, B. {Inhester}, and J. {Woch}:
  2005, `{Comparing magnetic field extrapolations with measurements of magnetic
  loops}'.
\newblock {\em A\&A} {\bf 433}, 701--705.

\bibitem[\protect\citeauthoryear{{Wiegelmann} and
  {Neukirch}}{2003}]{wiegelmann:etal03}
{Wiegelmann}, T. and T. {Neukirch}: 2003, `{Computing nonlinear force free
  coronal magnetic fields}'.
\newblock {\em Nonlinear Processes in Geophysics} {\bf 10}, 313--322.

\bibitem[\protect\citeauthoryear{{Wu} et~al.}{1990}]{wu:etal90}
{Wu}, S.~T., M.~T. {Sun}, H.~M. {Chang}, M.~J. {Hagyard}, and G.~A. {Gary}:
  1990, `{On the numerical computation of nonlinear force-free magnetic
  fields}'.
\newblock {\em ApJ} {\bf 362}, 698--708.

\bibitem[\protect\citeauthoryear{{Yan} and {Sakurai}}{2000}]{yan:etal00}
{Yan}, Y. and T. {Sakurai}: 2000, `{New Boundary Integral Equation
  Representation for Finite Energy Force-Free Magnetic Fields in Open Space
  above the Sun}'.
\newblock {\em Sol. Phys.} {\bf 195}, 89--109.

\end{thebibliography}

\end{article}
\end{document}